\documentclass[twocolumn,prd]{revtex4}

\usepackage{graphicx}% Include figure files
\usepackage{dcolumn}% Align table columns on decimal point
\usepackage{bm}% bold math
\usepackage{nicefrac}

\newcommand{\h}{\textrm{H}}
\newcommand{\D}{\ensuremath{d}}
\newcommand{\HeThree}{\mbox{$^3$He}}
\newcommand{\HeFour}{\mbox{$^4$He}}
\newcommand{\HeFive}{\mbox{$^5$He}}
\newcommand{\LiFive}{\mbox{$^5$Li}}
\newcommand{\LiSix}{\mbox{$^6$Li}}
\newcommand{\LiSeven}{\mbox{$^7$Li}}
\newcommand{\BeSeven}{\mbox{$^7$Be}}
\newcommand{\BeEight}{\mbox{$^8$Be}}
\newcommand{\BeNine}{\mbox{$^9$Be}}
\newcommand{\BK}{\mbox{$\E{9}$\,K}}
\newcommand{\MeV}{\text{MeV}}
\newcolumntype{d}{D{.}{.}{-1}}    %Align on decimal point
\newcolumntype{b}{D{(}{\ (}{-1}}  %Align on opening bracket of errors; put space 

\newcommand{\E}[1]{\ensuremath{\times 10^{#1}}}
\newcommand{\rtw}{\ensuremath{\rightarrow}}

\newcommand{\Cite}[1]{\mbox{Ref.~\cite{#1}}}
\newcommand{\Cites}[1]{\mbox{Refs.~\cite{#1}}}
\newcommand{\eref}[1]{(\ref{#1})}

\newcommand{\Tref}[1]{Table~\ref{#1}}
\newcommand{\Fig}[1]{Fig.~\ref{#1}}

\newcommand{\Sec}[1]{Section~\ref{#1}}

\begin{document}

\title{Effect of quark-mass variation on big bang nucleosynthesis}

\author{J. C. Berengut}
\affiliation{School of Physics, University of New South Wales, Sydney 2052, Australia}
\author{V. F. Dmitriev}
\affiliation{Budker Institute of Nuclear Physics, 630090, Novosibirsk-90, Russia}
\author{V. V. Flambaum}
\affiliation{School of Physics, University of New South Wales, Sydney 2052, Australia}

\date{14 July 2009}

\begin{abstract}

We calculate the effect of variation in the light-current quark mass, $m_q$, on standard big bang nucleosynthesis. A change in $m_q$ at during the era of nucleosynthesis affects nuclear reaction rates, and hence primordial abundances, via changes the binding energies of light nuclei. It is found that a relative variation of $\delta m_q/m_q = 0.016 \pm 0.005$ provides better agreement between observed primordial abundances and those predicted by theory. This is largely due to resolution of the existing discrepancies for \LiSeven. However this method ignores possible changes in the position of resonances in nuclear reactions. The predicted \LiSeven\ abundance has a strong dependence on the cross-section of the resonant reactions $\HeThree\,(\D, p)\,\HeFour$ and $t\,(\D,n)\,\HeFour$. We show that changes in $m_q$ at the time of BBN could shift the position of these resonances away from the Gamow window and lead to an increased production of \LiSeven, exacerbating the lithium problem.

\end{abstract}


\maketitle

\section{Introduction}

Measurements of the primordial baryon-to-photon ratio $\eta$ from the cosmic microwave background from WMAP \cite{dunkley09ajss}, coupled with precise measurements of the neutron half-life \cite{amsler08plb}, have made big bang nucleosynthesis (BBN) an essentially parameter-free theory \cite{amsler08plb,fields06npa,cyburt08arxiv}. In this paradigm excellent agreement has been obtained between predicted and observed abundances of deuterium and \HeFour\ (see, e.g. the Particle Data Group review~\cite{amsler08plb} and references therin).
However there is some disagreement for \LiSeven, the only other element for which the abundance has been measured to an accuracy at which fruitful comparison with theory can be made.
%(Extrapolation of \HeThree\ observations to primordial abundances are subject to large uncertainties.)
While the ``lithium problem'' has been known for some time, it has been exacerbated by recent measurements of the $\HeThree(\alpha,\gamma)\BeSeven$ reaction \cite{cyburt08prc}. Standard BBN theory with $\eta$ provided by WMAP~5 overproduces \LiSeven\ by a factor of 2.4 -- 4.3 (around $4$ -- $5 \sigma$) \cite{cyburt08arxiv}.

One possible solution to the lithium problem is that the physical constants of the early Universe may have been slightly different. In fact, such variations in the physical laws can be well-motivated theoretically in an expanding Universe; see \cite{uzan03rmp} for a review.
\Cite{dmitriev04prd} considered variation of the deuterium binding energy $B_d$ during primordial nucleosynthesis. BBN has a high sensitivity to $B_d$ since its value determines the temperature at which deuterium can withstand photo-disintegration and hence the time at which nucleosynthesis begins. Their best-fit result $\Delta B_d/B_d = -0.019 \pm 0.005$ resolved then-extant discrepancies between theory and observation in both \LiSeven\ and $\eta$ (or alternatively, in \LiSeven\ and \HeFour\ with $\eta$ fixed by WMAP).

More recently, \Cite{dent07prd} examined the response of BBN to variation of several physical parameters, including binding energies, in a linear approximation. These were coupled with calculated dependences of binding energies on $m_q$ in \cite{flambaum07prc}, which found that the \LiSeven\ abundance discrepancy could be resolved by a variation in light-quark mass of $\delta m_q/m_q = 0.013 \pm 0.002$. Crucially, the \HeFour\ and $d$ abundances were found to be relatively insensitive to $m_q$ and so the existing agreement between theory and observation in these elements was maintained.

In this paper we re-examine the dependence of light-element production on variation of the dimensionless parameter $X_q = m_q/\Lambda_{QCD}$ where $m_q$ is the light-quark mass and $\Lambda_{QCD}$ is the pole in the running strong-coupling constant. We follow \cite{flambaum07prc} and assume that $\Lambda_{QCD}$ is constant, calculating the dependence on the small parameter $m_q$. This is not an approximation. Rather it only means that we measure all dimensions ($m_q$, cross sections, etc) in units of $\Lambda_{QCD}$. Therefore $\delta m_q/m_q$ should be understood as $\delta X_q/X_q$. We take into account several effects that were not previously considered, most importantly the nonlinear dependence on $m_q$ and variation of resonance positions.

\section{Variation of binding energies}
\label{sec:binding_energies}

The energy released in each reaction, $Q$, is determined by the masses of the reactants and products, which in turn are determined by the nuclear binding energies. As noted in \cite{dent07prd}, the $Q$-values affect the forward (exothermic) reaction rates via phase space and radiative emission factors. For radiative capture reactions at low energy $E$ the $Q$-dependence is
\begin{equation}
\label{eq:E1_capture}
\sigma (E) \propto E_\gamma^3 \sim (Q+E)^3 \ .
\end{equation}
For low-energy reactions with two nucleons in the exit channel the dependence is proportional to the outgoing channel velocity, $v \sim (Q+E)^{1/2}$. When the outgoing particles are charged, the Gamow factor of the exit channel can also contribute:
\begin{equation}
\label{eq:sigma_transfer}
\sigma (E)\sim (Q+E)^{1/2} e^{-\sqrt{E_g/(Q+E)}} \ .
\end{equation}
The Gamow factor appears because of the Coulomb barrier to the reaction; $E_g=2 \pi^2 Z_1^2 Z_2^2 \alpha^2 \mu c^2$ where $\alpha$ is the fine-structure constant, $Z_1$ and $Z_2$ are the charge numbers of the products, and $\mu$ is the reduced mass of the products.
At BBN temperatures we can usually assume that $E \ll Q$. Expanding in $Q$,
\begin{equation}
\label{eq:sigma_transfer_linear}
\sigma = \sigma_0 \left[ 1 + \frac{1}{2}\left( 1+ \sqrt{\frac{E_g}{Q}}\right)
		    \frac{\delta Q}{Q} + ... \right]
\end{equation}
and we see that the Gamow term in \eref{eq:sigma_transfer} is generally small (it was neglected in \cite{dent07prd}). However it can be important for some reactions, for example in $\BeSeven\,(n,p)\,\LiSeven$, \mbox{$\sqrt{E_g/Q} = 2.17$}, i.e. it triples the effect of $\delta Q$ on the reaction rate.
%for example in $\HeThree\,(n,p)\,t$, \mbox{$\sqrt{E_g/Q} = 0.98$}, i.e. it doubles the effect of $\delta Q$ on the reaction rate.

The reverse reaction rates are simply related to the forward rates via statistical factors. From detailed balance one finds
\begin{equation}
\frac{\left< \sigma v \right>_\textrm{rev}}{\left< \sigma v \right>_\textrm{fwd}}
\sim e^{\nicefrac{-Q}{T}}
\end{equation}
and we see that the reverse reactions also provide sensitivity to $Q$.

An exception to the rule \eref{eq:E1_capture} is found in the reaction $p\,(n,\gamma)\,\D$, an important reaction because \D\ is a precursor to all further nucleosynthesis. This reaction is sensitive not only to $Q$ but also to the position of the virtual level with energy $\epsilon_\nu = 0.07$~\MeV. The sensitivity of this reaction to $Q$ was calculated in \cite{dmitriev04prd}
\begin{equation}
\label{eq:npdg}
\left< \sigma v \right> \sim \left[ 1 + \left( 5/2 + \sqrt{\frac{Q}{\epsilon_\nu}}\right) \frac{\delta Q}{Q} \right]\ .
\end{equation}
Note that \cite{dent07prd,flambaum07prc} did not take variation of the virtual level into account. In \Tref{tab:Bd} we show the linear dependence of abundances on the deuterium binding energy with different theories of variation. It shows the effect of variation of the virtual level, as well as the effect of including $B_\D$ variation on other $Q$-values and reaction rates.

\begin{table}
\caption{\label{tab:Bd}
$\partial\ln Y_a/\partial\ln B_D$, the dependence of nuclear abundances, $Y_a$, on deuterium binding energy under different assumptions: \\
1.\ Variation of virtual level not considered, $\left< \sigma v \right> \sim Q^{5/2}$. $Q$ changed only for $p\,(n,\gamma)\,\D$. \\
2.\ Variation of virtual level not considered; effect of $B_\D$ included in all reactions (similar to theory of \cite{dent07prd}). \\
3.\ $p\,(n,\gamma)\,\D$ changed according to \eref{eq:npdg}, including variation of the virtual level; effect of $B_\D$ on other reactions ignored (similar to theory of \cite{dmitriev04prd}). \\
4.\ $p\,(n,\gamma)\,\D$ changed according to \eref{eq:npdg}; effect of $B_\D$ included in all reactions. \\
}
\begin{tabular}{lddddd}
\hline \hline
Method & \D & \HeThree & \HeFour & \LiSix & \LiSeven \\
\hline
1.   & -4.04 & -1.75 & 0.68 & -3.17 & 10.59 \\
2.   & -2.91 & -2.08 & 0.67 & -6.58 & 9.41 \\
3.   & -5.12 & -1.29 & 0.70 & -4.23 & 17.99 \\
4.   & -4.00 & -1.62 & 0.69 & -7.64 & 16.81 \\
\hline\hline
\end{tabular}
\end{table}

We denote the sensitivity of nuclear binding energies to the light-current quark mass $m_q$ by
\begin{equation}
\label{eq:K}
K = \frac{\delta E / E}{\delta m_q / m_q}\ .
\end{equation}
Values of $K$ for several light nuclei were presented in \Cites{flambaum07prc,flambaum09prc}. We use the ``best values'' from these papers, given by the AV18+UIX nuclear Hamiltonians, with hadron mass variations calculated in terms of the $m_q$ using the Dyson-Schwinger equation calculation of \cite{flambaum06fbs}. From these one calculates the $m_q$-dependence of the $Q$ values, and therefore the reaction rates, and therefore the primordial abundances of light elements in BBN.

\begin{figure}[tb]
\includegraphics[width=0.45\textwidth]{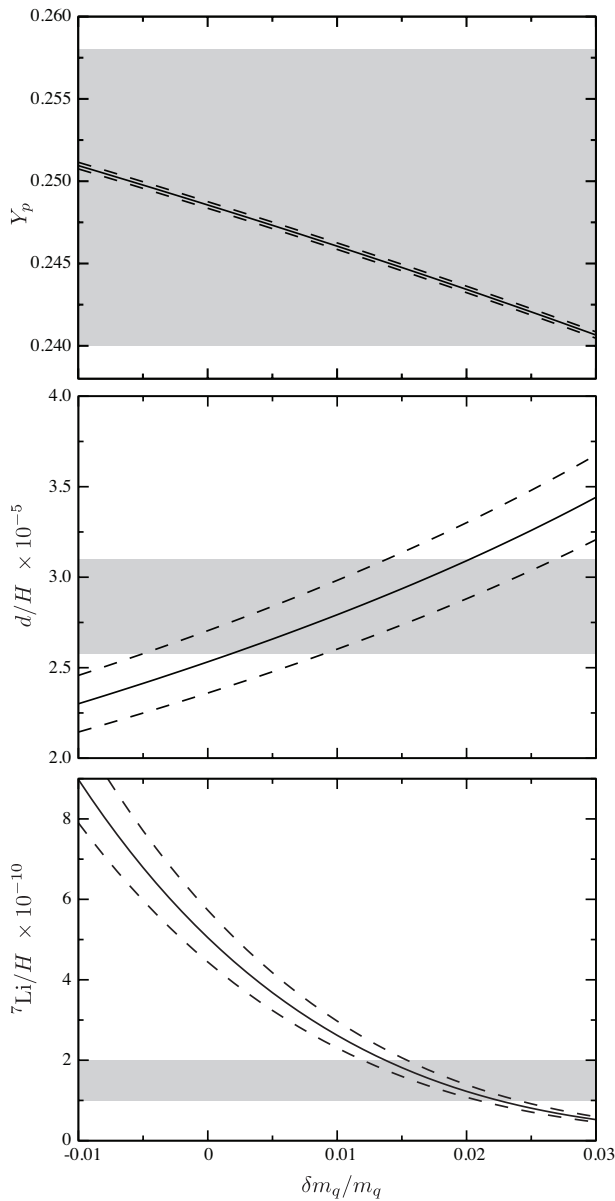}
\caption{\label{fig:All} Calculated \HeFour, \D, and \LiSeven\ abundances vs. relative change in light quark mass $m_q/\Lambda_{QCD}$ (solid lines). The ranges showed by the dashed lines are $1\sigma$~errors in the theory, assuming the relative errors are constant (i.e. these do not take into account any error in the $K$ factors of Eq.~\ref{eq:K}). The shaded areas show $1\sigma$~ranges of observed abundances (details in Appendix~\ref{sec:obs}).}
\end{figure}

In \Fig{fig:All} we present our predicted values of \HeFour, \D, and \LiSeven\ with different values of light-quark mass. Details of the calculations and explanation of observational abundances are presented in the appendices.
Comparing the observed and predicted abundances from the figures we obtain for \HeFour, \D, and \LiSeven\ respectively, \mbox{$\delta m_q/m_q = -0.002 \pm 0.037$}, \mbox{$0.012 \pm 0.011$}, and \mbox{$0.018 \pm 0.006$}. The three data sets are therefore consistent, with weighted mean
\begin{equation}
\label{eq:dmq_value}
\delta m_q/m_q = 0.016 \pm 0.005\ .
\end{equation}
It is seen that the \HeFour\ abundance has a low sensitivity to $m_q$; furthermore we show in \Fig{fig:All} the conservative observational error bounds provided by \cite{olive04apj}. Therefore, it is worth pointing out that the more tightly constrained abundance, $Y_p = 0.2477 \pm 0.0029$~\cite{peimbert07apj}, is also consistent with the variation \eref{eq:dmq_value}. It is clear from \Fig{fig:All} that taking into account the nonlinear dependence of BBN abundances on $m_q$ is important, particularly for \LiSeven. In fact, if we assume a linear response, as was done in \cite{dent07prd,flambaum07prc}, we instead obtain $\delta m_q/m_q = 0.014 \pm 0.002$.

As noted in \cite{flambaum07prc}, to take into account uncertainties in the theoretically derived quantities $K$ (Equation \ref{eq:K}) the final result \eref{eq:dmq_value} should be interpreted as \mbox{ $\delta m_q/m_q = k \cdot (0.016 \pm 0.005)$} where $k \sim 1$ and the accuracy in $k$ is approximately a factor of two.

\section{Resonances}

Of the most important reactions in BBN, the mirror reactions
\begin{eqnarray}
\HeThree\, (d, p)\, \HeFour \quad &\textrm{(Reaction 1)}& \nonumber \\
t\, (d, n)\, \HeFour \quad &\textrm{(Reaction 2)}& \nonumber
\end{eqnarray}
are the only reactions where the cross-section is dominated by a fairly narrow resonance. Therefore, one can hope for sensitivity of primordial abundances to the position of these resonances. (Note that the reaction $\BeSeven\, (n, p)\, \LiSeven$ is also dominated by a near-threshold resonance, however in this case the resonance is a rather broad and hence strong sensitivity can hardly be expected.)

Both of these reactions have the cross-sections with the general form
\begin{equation}
\sigma(E) = \frac{e^{-\sqrt{E_g/E}}}{E} \frac{P(E)}{(E-E_r)^2 + \Gamma_r^2/4}
\end{equation}
where $E_g$ is the Gamow energy of the reactants, $E_r$ and $\Gamma_r$ are resonance parameters, and $P(E)$ is a polynomial chosen to fit the measured reaction cross-section. In this work we use the cross-section fits of \Cite{cyburt04prd}, which give $E_r^{(1)} = 0.183$~\MeV, $\Gamma_r^{(1)} = 0.256$~\MeV\ and $E_r^{(2)} = 0.0482$~\MeV, $\Gamma_r^{(2)} = 0.0806$~\MeV\ for reactions 1 and 2, respectively.

Consider modification of the resonance positions, $E_r \rtw E_r + \delta E_r$, due to a variation of the fundamental constant $m_q$. Reaction 1 will be affected in the following way. The resonance is an excited state of \LiFive; that is, a compound nucleus with three protons and two neutrons: we call this state $\LiFive^{*}$. Similarly there is a state $\HeFive^{*}$ for reaction 2. Then
\begin{eqnarray}
\label{eq:Er1}
E_r^{(1)} &=& E_{^5\textrm{Li}^{*}} - E_{^3\textrm{He}} - E_d \\
\label{eq:Er2}
E_r^{(2)} &=& E_{^5\textrm{He}^{*}} - E_t - E_d
\end{eqnarray}
and so $E_{^5\textrm{Li}^{*}} = -9.76$~\MeV\ and $E_{^5\textrm{He}^{*}} = -10.66$~\MeV.
The change in the resonance position due to a variation in $m_q$ is therefore
\begin{eqnarray}
\delta E_r^{(1)} &=& \delta E_{^5\textrm{Li}^{*}} - \delta E_{^3\textrm{He}} - \delta E_d \\
 &=& \left( K_{^5\textrm{Li}^{*}} E_{^5\textrm{Li}^{*}} - K_{^3\textrm{He}} E_{^3\textrm{He}} - K_d E_d \right)
     \frac{\delta m_q}{m_q}
\end{eqnarray}
with the $K$ defined by \eref{eq:K}.

\begin{figure}[tb]
\includegraphics[width=0.45\textwidth]{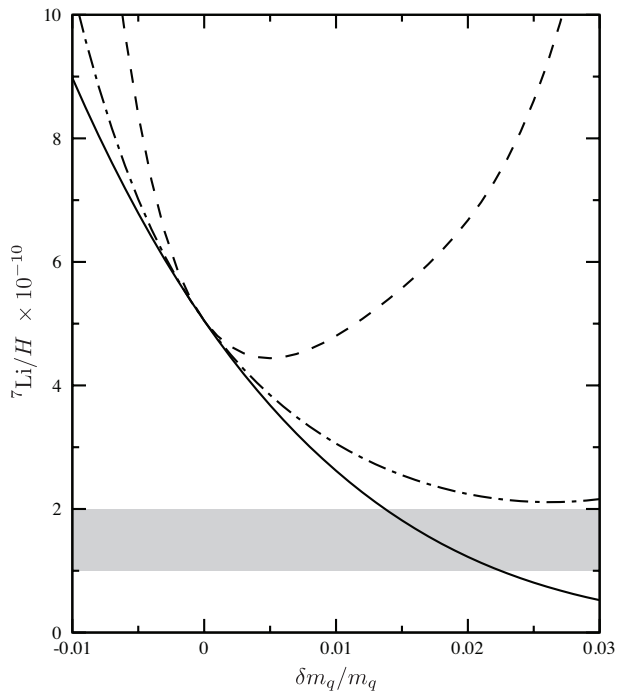}
\caption{\label{fig:LiSeven} Calculated \LiSeven\ abundance vs. relative change in light quark mass $m_q/\Lambda_{QCD}$. Solid line: no shifts in resonance positions included (same as solid line in \Fig{fig:All}); dashed line: resonance shifts according to assumption that resonant state varies as much as the ground state (equations~\ref{eq:K1} and \ref{eq:K2}); dot-dashed line: an averaged value of resonance-position sensitivity used. The shaded area shows the $1\sigma$ range of abundances.}
\end{figure}

Changes to the cross-section of reaction 1 affects the primordial abundances of \HeThree\ and \BeSeven, while changes in reaction 2 affect abundances of t and \LiSeven. Since t and \HeThree\ are not well constrained observationally, we choose to focus on \LiSeven. In \Fig{fig:LiSeven} we present \LiSeven\ abundance against variation of light quark mass $\delta m_q/m_q$ at $\eta = 6.23\E{-10}$, the WMAP5 value. For such a value of $\eta$, the majority of \LiSeven\ is created as \BeSeven\ (which $\beta$-captures to \LiSeven) via the reaction $\HeThree\,(\HeFour, \gamma)\,\BeSeven$.

We need to find $K_{^5\textrm{Li}^{*}}$ (and similarly $K_{^5\textrm{He}^{*}}$). One assumption is that the mass-energy of the resonance varies with the mass-energy in the incoming channel \cite{dent07prd}; in this case the resonance does not shift. This assumption corresponds to $K_{^5\textrm{Li}^{*}} = -1.54$ and $K_{^5\textrm{He}^{*}} = -1.44$. It corresponds to the solid line in \Fig{fig:LiSeven}.

A more reasonable guess is to assume that the variation of the resonant state $\LiFive^{*}$ will be approximately the same as that of the ground state \LiFive. This can be seen by considering the resonance and the ground state configurations as residing in the same potential.
The sensitivity of the ground state \HeFive\ to $m_q$ has been calculated $K_{^5\textrm{He}} = -1.24$~\cite{flambaum09prc}; $K_{^5\textrm{Li}}$ was not calculated explicitly, but its value will be very close to that of $\HeFive$. Our assumption of equal variation of the ground and excited state then gives
\begin{eqnarray}
\label{eq:K1}
K_{^5\textrm{Li}^{*}} &=& -3.35 \\
\label{eq:K2}
K_{^5\textrm{He}^{*}} &=& -3.19
\end{eqnarray}
This assumption corresponds to the dashed line in \Fig{fig:LiSeven}.

The equal-variation assumption in the previous paragraph represents an upper limit on the relationship between the ground and excited state. In reality the potential-dependence of the states may be different, in which case the shift of the \LiFive\ (or \HeFive) resonance may be smaller than the shift of the ground state. On the other hand a minimum value of $K$ for the resonance states is that of the ground state, $K_{^5\textrm{Li}^{*}} = K_{^5\textrm{Li}} = -1.24$. A reasonable, conservative, estimate is to take the average of these extremal values: $K_{^5\textrm{Li}^{*}} = -2.29$ and $K_{^5\textrm{He}^{*}} = -2.21$; this is the dot-dashed line in \Fig{fig:LiSeven}. Ultimately however, we require a nuclear calculation of sensitivity, of the kind presented in \Cites{flambaum07prc,flambaum09prc}.

%In the real-world case the excited state may not vary quite as strongly as the ground state. For example, consider a single-particle in a square well; while the energy of deeply bound levels vary linearly with the potential, a shallow level varies as the square root of the potential. Therefore, the shift of the \LiFive\ (or \HeFive) resonance may be smaller than the shift of the ground state. Furthermore, the effective potentials of the ground state and resonant state will not be exactly the same. The smallest possible value of $K$ for the resonance states is that of the ground state, $K_{^5\textrm{He}^{*}} = K_{^5\textrm{He}} = -1.24$. The largest value was calculated in the preceding paragraph (equations~\ref{eq:K1} and~\ref{eq:K2}). We therefore also present the response of BBN to variation in the resonance positions when $K$ takes the average of these extremal values: $K_{^5\textrm{Li}^{*}} = -2.29$ and $K_{^5\textrm{He}^{*}} = -2.21$; this is the dot-dashed line in \Fig{fig:LiSeven}.

The effect of $\delta E_r^{(1)}$ on BBN can be understood in the following way.
When the cross-section is convolved with a Maxwellian distribution, the exponential term gives rise to the ``Gamow window'' at energy
$E_0/E_g = (kT/2E_g)^{2/3}$. This reaction is most active at $kT \approx 0.07$ \MeV, at which time the Gamow window is at $E_0 = 0.180$~\MeV. This is remarkably close to the resonance energy for this reaction $E_r = 0.183$ \MeV. Therefore movement of the resonance position in either direction will reduce the cross-section for this reaction at the relevant temperatures. In turn this reduces the amount of \HeThree\ that is destroyed via reaction 1, leaving more to react with \HeFour\ to produce \BeSeven. On the other hand the effect of this reaction on \D\ and \HeFour\ abundances is minimal.

The effect of $\delta E_r^{(2)}$ is very similar: it reduces the amount of t destroyed in reaction 2, leaving more tritium to react with \HeFour\ to produce \LiSeven\ directly. Despite this production channel being suppressed at high $\eta$, the effect of $\delta E_r^{(2)}$ is still important for \LiSeven\ production because the relative effect of the variation is larger: $\delta E_r^{(2)}/\Gamma_r^{(2)} > \delta E_r^{(1)}/\Gamma_r^{(1)}$. The trends seen in \Fig{fig:LiSeven} are the same even at low $\eta$ since both reaction pathways behave in much the same way to variation in $m_q$.

\begin{figure}[tb]
\includegraphics[width=0.45\textwidth]{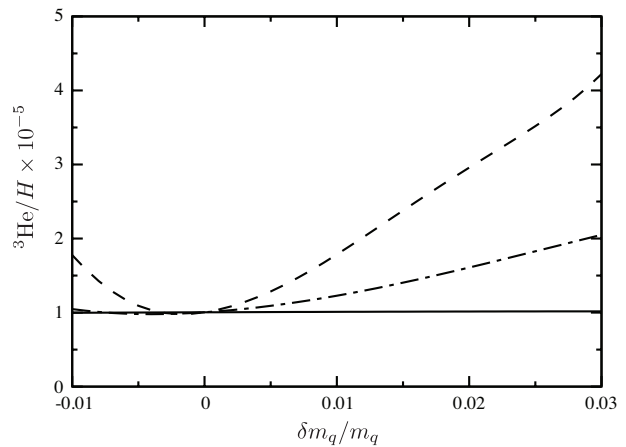}
\caption{\label{fig:HeThree} Calculated \HeThree\ abundance vs. relative change in light quark mass $m_q/\Lambda_{QCD}$. Solid line: no shifts in resonance positions included; dashed line: resonance shifts according to assumption that resonant state varies as much as the ground state (equations~\ref{eq:K1} and \ref{eq:K2}); dot-dashed line: an averaged value of resonance-position sensitivity used.}
\end{figure}

From \Fig{fig:LiSeven} we see that taking shifts in the resonance positions into account can destroy the agreement between theory and observation previously obtained by varying $m_q$. In the case where the shifts in the ground and resonant states vary by the same amount (dashed line), the \LiSeven\ discrepancy actually gets worse with variation in light quark mass. On the other hand the milder ``averaged $K$'' response (dot-dashed line) still significantly challenges the conclusions of \Sec{sec:binding_energies}. It is not appropriate to directly compare primordial \HeThree\ abundances with observations because of the complexity of the stellar evolution of this isotope \cite{vangioni-flam03apj}, however we note that primordial \HeThree\ production could also be greatly increased by movement of these resonances (\Fig{fig:HeThree}).

\section{Conclusion}

We have shown in \Sec{sec:binding_energies} that a variation in the light-quark mass during the era of big-bang nucleosynthesis of $\delta m_q/m_q = 0.016 \pm 0.005$ provides better agreement between theory and the observed primordial abundances. This is largely because it resolves the existing disagreement in \LiSeven\ abundances~\cite{cyburt08arxiv}.

However this conclusion is threatened when movement of the resonance positions in the reactions $\HeThree\, (d, p)\, \HeFour$ and $t\, (d, n)\, \HeFour$ is taken into account. These reactions strongly affect \LiSeven\ production during BBN; furthermore they are already ``on resonance'' meaning that movement of the resonance position in either direction increases \LiSeven. Our estimates suggest that the $\HeFive^{*}$ and $\LiFive^{*}$ resonances may be very sensitive to variation of $m_q/\Lambda_{QCD}$. Therefore it is very important that the sensitivity of these resonances to fundamental constants be studied in more detail using nuclear models.

This work is supported by the Australian Research Council. JCB thanks Daniel Wolf Savin for hospitality during the early stages of this work.

\appendix
\section{Observational abundances}
\label{sec:obs}

Our observational abundances largely follow the recommendations of the Particle Data Group review \cite{amsler08plb}. The deuterium abundances are derived from several studies of isotope-shifted Ly-$\alpha$ spectra in quasar absorption systems. Combined these give
\[
\D/H = (2.84 \pm 0.26) \E{-5}
\]
where the errors have been increased to account for the scatter between different systems.

\HeFour\ is observed in H II regions of low-metallicity dwarf galaxies. A very conservative estimate of observed \HeFour\ abundance comes from \cite{olive04apj}
\[
Y_p = 0.249 \pm 0.009 \ .
\]
The error here is significantly larger than other extrapolations to zero metallicity, e.g.~\cite{izotov07apj,peimbert07apj}.

Primordial \LiSeven\ abundance is determined from metal-poor Pop II stars in our galaxy. Lithium abundance does not vary over many orders of magnitude of metallicity in such stars; this is the Spite plateau \cite{spite82nat}.
Recent studies give abundances of
$(1.1-1.2 \pm 0.1)\E{-10}$~\cite{hosford09aap},
$(1.26 \pm 0.26)\E{-10}$~\cite{bonifacio07aap}, and
$(1.1-1.5)\E{-10}$~\cite{asplund06apj}.
Significantly higher results were obtained with different methods of obtaining effective temperature of the stars \cite{melendez04apj} since the derived lithium abundance is very sensitive to temperature. However no evidence for high temperatures was found in the studies \cite{asplund06apj,hosford09aap}.

On the other hand measurements of \LiSeven\ abundance in the globular cluster NGC 6397 give values of
$(2.19 \pm 0.28)\E{-10}$~\cite{bonifacio02aap},
$(1.91 \pm 0.44)\E{-10}$~\cite{pasquini96aap}, and
$(1.69 \pm 0.27)\E{-10}$~\cite{thevenin01aap}. The M 92 globular cluster yields a value of $(2.29\pm0.94)\E{-10}$~\cite{bonifacio02aap0}.
For a more detailed review and discussion of systematics, see, e.g.~\cite{amsler08plb,hosford09aap,cyburt08arxiv}.

In this paper we use the conservative range of
\[
\LiSeven/H = (1.5 \pm 0.5) \E{-10}\ ,
\]
which was also adopted in \cite{dent07prd}, although we note that some of the studies listed above give ranges as high as $2.5\E{-10}$.

\section{Computer code, reaction rates, and theoretical uncertainties}
\label{sec:calc}

In this work we calculate BBN abundances using a modified Kawano code \cite{kawano92fermilab}, with updated reactions from the NACRE collaboration \cite{angulo99npa}. For the most important reactions we use the cross-section fits in \cite{cyburt04prd}. The exceptions are $p\,(n, \gamma)\D$ where we use the calculations of \cite{ando06prc}, and the recently measured reaction $\HeThree\,(\HeFour, \gamma)\,\BeSeven$ for which we use the fits provided in \cite{cyburt08prc}. We have not calculated errors in the theoretical prediction; instead we simply take the relative errors from the recent calculations of Cyburt, Fields and Olive \cite{cyburt08arxiv} which use a very similar reaction network and the latest physical data:
\begin{eqnarray}
Y_p &=& 0.2486 \pm 0.0002 \\
\D/\h &=& (2.49 \pm 0.17) \E{-5} \\
\LiSeven/\h &=& (5.24^{+0.71}_{-0.62}) \E{-10}
\end{eqnarray}
These values compare well with the results from our code of \mbox{$Y_p = 0.2486$}, \mbox{$\D/\h = 2.53\E{-5}$}, and \mbox{$\LiSeven/\h = 5.05\E{-10}$}.

\bibliography{references}

\begin{thebibliography}{28}
\expandafter\ifx\csname natexlab\endcsname\relax\def\natexlab#1{#1}\fi
\expandafter\ifx\csname bibnamefont\endcsname\relax
  \def\bibnamefont#1{#1}\fi
\expandafter\ifx\csname bibfnamefont\endcsname\relax
  \def\bibfnamefont#1{#1}\fi
\expandafter\ifx\csname citenamefont\endcsname\relax
  \def\citenamefont#1{#1}\fi
\expandafter\ifx\csname url\endcsname\relax
  \def\url#1{\texttt{#1}}\fi
\expandafter\ifx\csname urlprefix\endcsname\relax\def\urlprefix{URL }\fi
\providecommand{\bibinfo}[2]{#2}
\providecommand{\eprint}[2][]{\url{#2}}

\bibitem[{\citenamefont{\protect{Dunkley \etal}}(2009)}]{dunkley09ajss}
\bibinfo{author}{\bibfnamefont{J.}~\bibnamefont{\protect{Dunkley \etal}}},
  \bibinfo{journal}{\ajss} \textbf{\bibinfo{volume}{180}}, \bibinfo{pages}{306}
  (\bibinfo{year}{2009}).

\bibitem[{\citenamefont{\protect{Amsler \etal\ (Particle Data
  Group)}}(2008)}]{amsler08plb}
\bibinfo{author}{\bibfnamefont{C.}~\bibnamefont{\protect{Amsler \etal\
  (Particle Data Group)}}}, \bibinfo{journal}{\plb}
  \textbf{\bibinfo{volume}{667}}, \bibinfo{pages}{1} (\bibinfo{year}{2008}).

\bibitem[{\citenamefont{Fields and Olive}(2006)}]{fields06npa}
\bibinfo{author}{\bibfnamefont{B.~D.} \bibnamefont{Fields}} \bibnamefont{and}
  \bibinfo{author}{\bibfnamefont{K.~A.} \bibnamefont{Olive}},
  \bibinfo{journal}{\npa} \textbf{\bibinfo{volume}{777}}, \bibinfo{pages}{208}
  (\bibinfo{year}{2006}).

\bibitem[{\citenamefont{Cyburt et~al.}(2008)\citenamefont{Cyburt, Fields, and
  Olive}}]{cyburt08arxiv}
\bibinfo{author}{\bibfnamefont{R.~H.} \bibnamefont{Cyburt}},
  \bibinfo{author}{\bibfnamefont{B.~D.} \bibnamefont{Fields}},
  \bibnamefont{and} \bibinfo{author}{\bibfnamefont{K.~A.} \bibnamefont{Olive}}
  (\bibinfo{year}{2008}), \bibinfo{note}{arXiv:0808.2818}.

\bibitem[{\citenamefont{Cyburt and Davids}(2008)}]{cyburt08prc}
\bibinfo{author}{\bibfnamefont{R.~H.} \bibnamefont{Cyburt}} \bibnamefont{and}
  \bibinfo{author}{\bibfnamefont{B.}~\bibnamefont{Davids}},
  \bibinfo{journal}{\prc} \textbf{\bibinfo{volume}{78}},
  \bibinfo{pages}{064614} (\bibinfo{year}{2008}).

\bibitem[{\citenamefont{Uzan}(2003)}]{uzan03rmp}
\bibinfo{author}{\bibfnamefont{J.-P.} \bibnamefont{Uzan}},
  \bibinfo{journal}{\rmp} \textbf{\bibinfo{volume}{75}}, \bibinfo{pages}{403}
  (\bibinfo{year}{2003}).

\bibitem[{\citenamefont{Dmitriev et~al.}(2004)\citenamefont{Dmitriev, Flambaum,
  and Webb}}]{dmitriev04prd}
\bibinfo{author}{\bibfnamefont{V.~F.} \bibnamefont{Dmitriev}},
  \bibinfo{author}{\bibfnamefont{V.~V.} \bibnamefont{Flambaum}},
  \bibnamefont{and} \bibinfo{author}{\bibfnamefont{J.~K.} \bibnamefont{Webb}},
  \bibinfo{journal}{\prd} \textbf{\bibinfo{volume}{69}},
  \bibinfo{pages}{063506} (\bibinfo{year}{2004}).

\bibitem[{\citenamefont{Dent et~al.}(2007)\citenamefont{Dent, Stern, and
  Wetterich}}]{dent07prd}
\bibinfo{author}{\bibfnamefont{T.}~\bibnamefont{Dent}},
  \bibinfo{author}{\bibfnamefont{S.}~\bibnamefont{Stern}}, \bibnamefont{and}
  \bibinfo{author}{\bibfnamefont{C.}~\bibnamefont{Wetterich}},
  \bibinfo{journal}{\prd} \textbf{\bibinfo{volume}{76}},
  \bibinfo{pages}{063513} (\bibinfo{year}{2007}).

\bibitem[{\citenamefont{Flambaum and Wiringa}(2007)}]{flambaum07prc}
\bibinfo{author}{\bibfnamefont{V.~V.} \bibnamefont{Flambaum}} \bibnamefont{and}
  \bibinfo{author}{\bibfnamefont{R.~B.} \bibnamefont{Wiringa}},
  \bibinfo{journal}{\prc} \textbf{\bibinfo{volume}{76}},
  \bibinfo{pages}{054002} (\bibinfo{year}{2007}).

\bibitem[{\citenamefont{Flambaum and Wiringa}(2009)}]{flambaum09prc}
\bibinfo{author}{\bibfnamefont{V.~V.} \bibnamefont{Flambaum}} \bibnamefont{and}
  \bibinfo{author}{\bibfnamefont{R.~B.} \bibnamefont{Wiringa}},
  \bibinfo{journal}{\prc} \textbf{\bibinfo{volume}{79}},
  \bibinfo{pages}{034302} (\bibinfo{year}{2009}).

\bibitem[{\citenamefont{Flambaum et~al.}(2006)\citenamefont{Flambaum, H\"{o}ll,
  Jaikumar, Roberts, and Wright}}]{flambaum06fbs}
\bibinfo{author}{\bibfnamefont{V.~V.} \bibnamefont{Flambaum}},
  \bibinfo{author}{\bibfnamefont{A.}~\bibnamefont{H\"{o}ll}},
  \bibinfo{author}{\bibfnamefont{P.}~\bibnamefont{Jaikumar}},
  \bibinfo{author}{\bibfnamefont{C.~D.} \bibnamefont{Roberts}},
  \bibnamefont{and} \bibinfo{author}{\bibfnamefont{S.~V.}
  \bibnamefont{Wright}}, \bibinfo{journal}{\fbs} \textbf{\bibinfo{volume}{38}},
  \bibinfo{pages}{31} (\bibinfo{year}{2006}).

\bibitem[{\citenamefont{Olive and Skillman}(2004)}]{olive04apj}
\bibinfo{author}{\bibfnamefont{K.~A.} \bibnamefont{Olive}} \bibnamefont{and}
  \bibinfo{author}{\bibfnamefont{E.~D.} \bibnamefont{Skillman}},
  \bibinfo{journal}{\apj} \textbf{\bibinfo{volume}{617}}, \bibinfo{pages}{29}
  (\bibinfo{year}{2004}).

\bibitem[{\citenamefont{Peimbert et~al.}(2007)\citenamefont{Peimbert,
  Luridiana, and Peimbert}}]{peimbert07apj}
\bibinfo{author}{\bibfnamefont{M.}~\bibnamefont{Peimbert}},
  \bibinfo{author}{\bibfnamefont{V.}~\bibnamefont{Luridiana}},
  \bibnamefont{and} \bibinfo{author}{\bibfnamefont{A.}~\bibnamefont{Peimbert}},
  \bibinfo{journal}{\apj} \textbf{\bibinfo{volume}{666}}, \bibinfo{pages}{636}
  (\bibinfo{year}{2007}).

\bibitem[{\citenamefont{Cyburt}(2004)}]{cyburt04prd}
\bibinfo{author}{\bibfnamefont{R.~H.} \bibnamefont{Cyburt}},
  \bibinfo{journal}{\prd} \textbf{\bibinfo{volume}{70}},
  \bibinfo{pages}{023505} (\bibinfo{year}{2004}).

\bibitem[{\citenamefont{Vangioni-Flam et~al.}(2003)\citenamefont{Vangioni-Flam,
  Olive, Fields, and Cass\'e}}]{vangioni-flam03apj}
\bibinfo{author}{\bibfnamefont{E.}~\bibnamefont{Vangioni-Flam}},
  \bibinfo{author}{\bibfnamefont{K.~A.} \bibnamefont{Olive}},
  \bibinfo{author}{\bibfnamefont{B.~D.} \bibnamefont{Fields}},
  \bibnamefont{and} \bibinfo{author}{\bibfnamefont{M.}~\bibnamefont{Cass\'e}},
  \bibinfo{journal}{\apj} \textbf{\bibinfo{volume}{585}}, \bibinfo{pages}{611}
  (\bibinfo{year}{2003}).

\bibitem[{\citenamefont{Izotov et~al.}(2007)\citenamefont{Izotov, Thuan, and
  Stasi\'nska}}]{izotov07apj}
\bibinfo{author}{\bibfnamefont{Y.~I.} \bibnamefont{Izotov}},
  \bibinfo{author}{\bibfnamefont{T.~X.} \bibnamefont{Thuan}}, \bibnamefont{and}
  \bibinfo{author}{\bibfnamefont{G.}~\bibnamefont{Stasi\'nska}},
  \bibinfo{journal}{\apj} \textbf{\bibinfo{volume}{662}}, \bibinfo{pages}{15}
  (\bibinfo{year}{2007}).

\bibitem[{\citenamefont{Spite and Spite}(1982)}]{spite82nat}
\bibinfo{author}{\bibfnamefont{M.}~\bibnamefont{Spite}} \bibnamefont{and}
  \bibinfo{author}{\bibfnamefont{F.}~\bibnamefont{Spite}},
  \bibinfo{journal}{\nat} \textbf{\bibinfo{volume}{297}}, \bibinfo{pages}{483}
  (\bibinfo{year}{1982}).

\bibitem[{\citenamefont{Hosford et~al.}(2009)\citenamefont{Hosford, Ryan,
  P\'erez, Norris, and Olive}}]{hosford09aap}
\bibinfo{author}{\bibfnamefont{A.}~\bibnamefont{Hosford}},
  \bibinfo{author}{\bibfnamefont{S.~G.} \bibnamefont{Ryan}},
  \bibinfo{author}{\bibfnamefont{A.~E.~G.} \bibnamefont{P\'erez}},
  \bibinfo{author}{\bibfnamefont{J.~E.} \bibnamefont{Norris}},
  \bibnamefont{and} \bibinfo{author}{\bibfnamefont{K.~A.} \bibnamefont{Olive}},
  \bibinfo{journal}{\aap} \textbf{\bibinfo{volume}{493}}, \bibinfo{pages}{601}
  (\bibinfo{year}{2009}).

\bibitem[{\citenamefont{Bonifacio et~al.}(2007)\citenamefont{Bonifacio, Molaro,
  Sivarani, Cayrel, Spite, Spite, Plez, Andersen, Barbuy, Beers
  et~al.}}]{bonifacio07aap}
\bibinfo{author}{\bibfnamefont{P.}~\bibnamefont{Bonifacio}},
  \bibinfo{author}{\bibfnamefont{P.}~\bibnamefont{Molaro}},
  \bibinfo{author}{\bibfnamefont{T.}~\bibnamefont{Sivarani}},
  \bibinfo{author}{\bibfnamefont{R.}~\bibnamefont{Cayrel}},
  \bibinfo{author}{\bibfnamefont{M.}~\bibnamefont{Spite}},
  \bibinfo{author}{\bibfnamefont{F.}~\bibnamefont{Spite}},
  \bibinfo{author}{\bibfnamefont{B.}~\bibnamefont{Plez}},
  \bibinfo{author}{\bibfnamefont{J.}~\bibnamefont{Andersen}},
  \bibinfo{author}{\bibfnamefont{B.}~\bibnamefont{Barbuy}},
  \bibinfo{author}{\bibfnamefont{T.~C.} \bibnamefont{Beers}},
  \bibnamefont{et~al.}, \bibinfo{journal}{\aap} \textbf{\bibinfo{volume}{462}},
  \bibinfo{pages}{851} (\bibinfo{year}{2007}).

\bibitem[{\citenamefont{Asplund et~al.}(2006)\citenamefont{Asplund, Nissen,
  Lambert, Primas, and Smith}}]{asplund06apj}
\bibinfo{author}{\bibfnamefont{M.}~\bibnamefont{Asplund}},
  \bibinfo{author}{\bibfnamefont{P.~E.} \bibnamefont{Nissen}},
  \bibinfo{author}{\bibfnamefont{D.~L.} \bibnamefont{Lambert}},
  \bibinfo{author}{\bibfnamefont{F.}~\bibnamefont{Primas}}, \bibnamefont{and}
  \bibinfo{author}{\bibfnamefont{V.~V.} \bibnamefont{Smith}},
  \bibinfo{journal}{\apj} \textbf{\bibinfo{volume}{664}}, \bibinfo{pages}{229}
  (\bibinfo{year}{2006}).

\bibitem[{\citenamefont{Mel\'endez and Ram\'irez}(2004)}]{melendez04apj}
\bibinfo{author}{\bibfnamefont{J.}~\bibnamefont{Mel\'endez}} \bibnamefont{and}
  \bibinfo{author}{\bibfnamefont{I.}~\bibnamefont{Ram\'irez}},
  \bibinfo{journal}{\apj} \textbf{\bibinfo{volume}{615}}, \bibinfo{pages}{L33}
  (\bibinfo{year}{2004}).

\bibitem[{\citenamefont{Bonifacio et~al.}(2002)\citenamefont{Bonifacio,
  Pasquini, Spite, Bragaglia, Carretta, Castellani, Centuri\`on, Chieffi,
  Claudi, Clementini et~al.}}]{bonifacio02aap}
\bibinfo{author}{\bibfnamefont{P.}~\bibnamefont{Bonifacio}},
  \bibinfo{author}{\bibfnamefont{L.}~\bibnamefont{Pasquini}},
  \bibinfo{author}{\bibfnamefont{F.}~\bibnamefont{Spite}},
  \bibinfo{author}{\bibfnamefont{A.}~\bibnamefont{Bragaglia}},
  \bibinfo{author}{\bibfnamefont{E.}~\bibnamefont{Carretta}},
  \bibinfo{author}{\bibfnamefont{V.}~\bibnamefont{Castellani}},
  \bibinfo{author}{\bibfnamefont{M.}~\bibnamefont{Centuri\`on}},
  \bibinfo{author}{\bibfnamefont{A.}~\bibnamefont{Chieffi}},
  \bibinfo{author}{\bibfnamefont{R.}~\bibnamefont{Claudi}},
  \bibinfo{author}{\bibfnamefont{G.}~\bibnamefont{Clementini}},
  \bibnamefont{et~al.}, \bibinfo{journal}{\aap} \textbf{\bibinfo{volume}{390}},
  \bibinfo{pages}{91} (\bibinfo{year}{2002}).

\bibitem[{\citenamefont{Pasquini and Molaro}(1996)}]{pasquini96aap}
\bibinfo{author}{\bibfnamefont{L.}~\bibnamefont{Pasquini}} \bibnamefont{and}
  \bibinfo{author}{\bibfnamefont{P.}~\bibnamefont{Molaro}},
  \bibinfo{journal}{\aap} \textbf{\bibinfo{volume}{307}}, \bibinfo{pages}{761}
  (\bibinfo{year}{1996}).

\bibitem[{\citenamefont{Th\'evenin et~al.}(2001)\citenamefont{Th\'evenin,
  Charbonnel, de~Freitas~Pacheco, Idiart, Jasniewicz, de~Laverny, and
  Plez}}]{thevenin01aap}
\bibinfo{author}{\bibfnamefont{F.}~\bibnamefont{Th\'evenin}},
  \bibinfo{author}{\bibfnamefont{C.}~\bibnamefont{Charbonnel}},
  \bibinfo{author}{\bibfnamefont{J.~A.} \bibnamefont{de~Freitas~Pacheco}},
  \bibinfo{author}{\bibfnamefont{T.~P.} \bibnamefont{Idiart}},
  \bibinfo{author}{\bibfnamefont{G.}~\bibnamefont{Jasniewicz}},
  \bibinfo{author}{\bibfnamefont{P.}~\bibnamefont{de~Laverny}},
  \bibnamefont{and} \bibinfo{author}{\bibfnamefont{B.}~\bibnamefont{Plez}},
  \bibinfo{journal}{\aap} \textbf{\bibinfo{volume}{373}}, \bibinfo{pages}{905}
  (\bibinfo{year}{2001}).

\bibitem[{\citenamefont{Bonifacio}(2002)}]{bonifacio02aap0}
\bibinfo{author}{\bibfnamefont{P.}~\bibnamefont{Bonifacio}},
  \bibinfo{journal}{\aap} \textbf{\bibinfo{volume}{395}}, \bibinfo{pages}{515}
  (\bibinfo{year}{2002}).

\bibitem[{\citenamefont{Kawano}(1992)}]{kawano92fermilab}
\bibinfo{author}{\bibfnamefont{L.}~\bibnamefont{Kawano}}
  (\bibinfo{year}{1992}), \bibinfo{note}{preprint FERMILAB-Pub-92/04-A}.

\bibitem[{\citenamefont{Angulo et~al.}(1999)\citenamefont{Angulo, Arnould,
  Rayet, Descouvemont, Baye, Leclercq-Willain, Coc, Barhoumi, Aguer, Rolfs
  et~al.}}]{angulo99npa}
\bibinfo{author}{\bibfnamefont{C.}~\bibnamefont{Angulo}},
  \bibinfo{author}{\bibfnamefont{M.}~\bibnamefont{Arnould}},
  \bibinfo{author}{\bibfnamefont{M.}~\bibnamefont{Rayet}},
  \bibinfo{author}{\bibfnamefont{P.}~\bibnamefont{Descouvemont}},
  \bibinfo{author}{\bibfnamefont{D.}~\bibnamefont{Baye}},
  \bibinfo{author}{\bibfnamefont{C.}~\bibnamefont{Leclercq-Willain}},
  \bibinfo{author}{\bibfnamefont{A.}~\bibnamefont{Coc}},
  \bibinfo{author}{\bibfnamefont{S.}~\bibnamefont{Barhoumi}},
  \bibinfo{author}{\bibfnamefont{P.}~\bibnamefont{Aguer}},
  \bibinfo{author}{\bibfnamefont{C.}~\bibnamefont{Rolfs}},
  \bibnamefont{et~al.}, \bibinfo{journal}{\npa} \textbf{\bibinfo{volume}{656}},
  \bibinfo{pages}{3} (\bibinfo{year}{1999}).

\bibitem[{\citenamefont{Ando et~al.}(2006)\citenamefont{Ando, Cyburt, Hong, and
  Hyun}}]{ando06prc}
\bibinfo{author}{\bibfnamefont{S.}~\bibnamefont{Ando}},
  \bibinfo{author}{\bibfnamefont{R.~H.} \bibnamefont{Cyburt}},
  \bibinfo{author}{\bibfnamefont{S.~W.} \bibnamefont{Hong}}, \bibnamefont{and}
  \bibinfo{author}{\bibfnamefont{C.~H.} \bibnamefont{Hyun}},
  \bibinfo{journal}{\prc} \textbf{\bibinfo{volume}{74}},
  \bibinfo{pages}{025809} (\bibinfo{year}{2006}).

\end{thebibliography}

\end{document}